\documentclass[12pt]{iopart}
\newcommand{\ui}{\left( u_{i}+\frac{\partial}{\partial u_{i}^{*} } \right)}
\newcommand{\uj}{\left( u_{j}+\frac{\partial}{\partial u_{j}^{*} } \right)}

\newcommand{\uid}{\left( u_{i}^{*}+\frac{\partial}{\partial u_i } \right)}

\begin{document}
\title{Rayleigh-Schr\"{o}dinger-Goldstone variational perturbation
       theory for many fermion systems}
\author{Sang Koo You$^{1,2}$ and Chul Koo Kim$^1$ }
\address{1. Institute of Physics and Applied Physics, Yonsei University, Seoul 120-749, Korea}
\address{2. Department of Physics, University of Incheon, Incheon 402-749, Korea}

\begin{abstract}
We present a Rayleigh-Schr\"{o}dinger-Goldstone perturbation
formalism for many fermion systems. Based on this formalism,
variational perturbation scheme which goes beyond the Gaussian
approximation is developed. In order to go beyond the Gaussian
approximation, we identify a parent Hamiltonian which has an
effective Gaussian vacuum as a variational solution and carry out
further perturbation with respect to the renormalized interaction
using Goldstone's expansion. Perturbation rules for the ground
state wavefunctional and energy are found.  Useful commuting
relations between operators and the Gaussian wavefunctional are
also found, which could reduce the calculational efforts
substantially. As examples, we calculate the first order
correction to the Gaussian wavefunctional and the second order
correction to the ground state of an electron gas system with the
Yukawa-type interaction.

\end{abstract}

\maketitle

\section{Introduction}
Field theories can be constructed from three kinds of pictures
which are called Heisenberg, Interaction and Schr\"{o}dinger
pictures [1]. Among them, the Heisenberg picture is known to
provide a convenient basis for the study of dynamics of operators
and systematic perturbative improvement on physical quantities. On
the other hand, the Schr\"{o}dinger picture approach  focuses on
the dynamics of wavefunction, which allows detailed study on time
evolution. However, in this approach, it is known that systematic
improvement on obtained result is rather difficult. Instead, it
has a powerful technique which is called a variational method
which allows nonperturbative access to problems, so that it could
be applied to strongly correlated systems where perturbative
approaches break down.

There have been successful applications of the Schr\"{o}dinger
picture to field theories [2-8] and later on to nonrelativistic
many-particle systems [9-15]. While it has been proved useful,
still the controlable range of trial wavefunctionals has been
found to be narrow and further improvement from results of a trial
wavefunctional  demands much endeavor in field theories.
Therefore, recent investigations on the picture have focused on
overcoming its drawbacks towards a more managable theory.

Gaussian trial wavefunctional apporach in bose fields has been
widely applied and proven to be an efficient and powerful method
in the Schr\"{o}dinger picture [3,8]. Further improvements beyond
the Gaussian approxiamtion have been investigated mostly in two
directions. One is to try it with non-Gaussian wavefunctionals
[16,17], and the other is to perform  appropriate expansions based
on Gaussian trial wavefunctional [18-24].

In fermi fields, a convenient prescription for operators was
proposed by Floreanini and Jackiw [5]. It was shown that this
approach gives successful results on Gaussian approximations of
fermi fields as in bose fields. However, in contrast to the
bosonic case [1], no successful Rayleigh-Schr\"{o}dinger type
perturbation formalism for fermi fields has been proposed so far,
although several alternative schemes including functional
integrals [25] and background field methods [8] have been reported
using the Floreanini-Jackiw representation(FJR). Since the
Rayleigh-Schr\"{o}dinger perturbation formalism is the most
familiar form of perturbation theories and especially suitable for
the Sch\"{o}dinger picture representation in quantum mechanics, it
is rather puzzling that it is not so in field theories. Indeed,
the Rayleigh-Schr\"{o}dinger perturbation scheme for bose fields
was formulated early [1]. However, for fermi fields, Grassmann
nature of functional variables has not allowed no such
straighforward formalism as for bose fields.

In this paper, we present a successful perturbation formalism for
fermionic many-body systems combining the Rayleigh-Schr\"{o}dinger
perturbation with Goldstone's (RSG) expansion.  The scheme will be
used to formulate a variational perturbation scheme, which
provides a systematic improvement beyond the Gaussian
approximation. In order to make the presentation compact, we
directly formulate a RSG perturbation scheme based on the
variational Gaussian wavefunctionals. The formalism is trivially
reduced to the conventional perturbation scheme by using a true
Gaussian vacuum instead of an effective Gaussian vacuum. This
point will be made clear once the formalism is given. It will be
shown that the FJR provides surprisingly simple expressions for
fermion wavefunctionals. As examples, we calculate the first order
correction to the Gaussian wavefunctional and the second order
energy of an electron gas with the Yukawa-type interaction.

\section{Floreanini-Jackiw representation}
The action of fermi field operators on Hilbert space can be
represented by  product or derivatives of Grassmann variables in
the Grassmann function space. Then, Schr\"{o}dinger equations are
transformed into functional differential equations [1]. Among many
possible representations preserving the anticommuting relations of
fermi field operators, the simplest one is generated from the
fermion coherent states which are defined as eigenstates of
annihilation operators [26]. In this representation, annihilation
operators \{$a_i$\} are described by Grassmann variables
\{$u_i^*$\}, and creation opeators \{$a_i^\dagger$\} , by
Grassmann derivatives \{$\frac{\partial}{\partial u_i^*}$\}. A
representation without a complex conjugate notation $*$ is also
possible and has been used by Duncan et al. [4,9]. However, we
find that the representation proposed by Floreanini and Jackiw is
more convenient to formulate the present theory.

In the FJR, the creation and annihilation operators are given as
follows [5];
\begin{eqnarray}
a_{i}^{\dagger} = \frac{1}{\sqrt{2}} \uid  \ \ , \ \ \ a_{i} =
\frac{1}{\sqrt{2}} \ui ,
\end{eqnarray}
where $u_i$ and $u_i^*$ are Grassmann variables.
 In the FJR for free
fermi systems, vacuums are described by Gaussian wavefunctionals
just as for bosonic systems. However, the norms between basis
states for the FJR are not orthogonal since $\langle u'{u'}^{*} |u
u^* \rangle = e^{{{u}_{i}'}^* u_i - {u}_{i}^* {u}_{i}' }$. It can
be proven by the fact that four kinds of states relative to a
quantum number $i$ are possible. They are represented by $1$,
$u_{i}^*$, $u_i$ and $u_{i}^* u_i$. Actually, only two kinds of
physical states for $i$ exist. One is the occupied state by a
fermion and the other unoccupied. Thus, the FJR is a reducible
representation and only a subspace of the total Grassmann
functional space can be used to describe physically well-defined
wavefunctionals. The subspace can be constructed only after a
physical vacuum wavefunctional is chosen. The dual space
wavefunctional $\bar{\Psi}$ for the wavefunctional $\Psi$ is
defined as
\begin{eqnarray}
\bar{\Psi} =\int D[u'{u'}^{*}] \Psi e^{{{u}_{i}'}^* u_i -
{u}_{i}^* {u}_{i}' },
\end{eqnarray}
where the norm is calculated by $ \langle \Psi|\Psi \rangle = \int
D[u{u}^{*}]\bar{\Psi} \Psi$. It can be easily shown that the free
fermion vacuum is described by Gaussian wavefunctional [5] as
\begin{eqnarray}
\Psi_G &=& e^{G_{ij}u_{i}^{*}
u_j}, \\
\bar{\Psi}_G &=& {\rm Det} G^{\dagger} e^{{\bar{G} }_{ij}u_{i}^{*}
u_j},
\end{eqnarray}
where $\bar{G} = {G^{\dagger}}^{-1}$ and we find the dual
wavefunctional is also Gaussian. Gaussian expectation values of
some normal ordered operators are calculated as
\begin{eqnarray}
&&\frac{\langle \Psi_G |  :{\cal O}(a^\dagger , a): | \Psi_G
\rangle}{\langle \Psi_G | \Psi_G \rangle} \nonumber \\
&=& :{\cal O} \left( \frac{1}{\sqrt{2}} \frac{\partial}{\partial
J_b } (I+\bar{G})_{bi} , \frac{1}{\sqrt{2}}
 (I+G)_{ja}\frac{\partial}{\partial J_a^* } \right): e^{(G+\bar{G})_{ij}^{-1}J_i^* J_j}|_{J,J^* =
 0},
\end{eqnarray}
where $I$ is an identity matrix and $J, J^*$ are source fields
which are inserted into the Gaussian wavefunctional as
$e^{G_{ij}u_{i}^{*} u_j +J_i^* u_i -u_i^* J_i}$ during the
calculation. It allows that the Grassmann integrals are
represented by Grassmann derivatives of source fields.

For the free field Hamiltonian, $H_0 = h_{ij} a_{i}^{\dagger}a_j$,
the functional Schr\"{o}dinger equation in the FJR is
\begin{eqnarray}
\frac{1}{2}h_{ij} \uid \uj \Psi =E_0 \Psi .
\end{eqnarray}
This equation contains second order derivatives and a quadratic
term which are similar to the harmonic oscillator problem. Thus,
we expect the Gaussian wavefunctional, Eq.(3), as a vacuum. In
order for Eq.(3) to be an eigenfunctional with an eigenvalue $E_0
= \frac{1}{2} {\rm Tr} h (I+G)$ in Eq.(6), $G$ should satisfy the
following condition;
\begin{eqnarray}
(I-G) h (I+G)=0.
\end{eqnarray}
Eq.(7) has trivial and non-trivial solutions. Trivial solutions
are $G= \pm I$ and non-trivial ones, $G=\pm \frac{h}{\sqrt{h^2}}$.
Here, we note that the number of particles $N$ is given by
$N=\frac{1}{2} {\rm Tr}(I+G)$. Therefore, $G$ can be expressed as
$G=- \frac{h-\mu I}{\sqrt{(h-\mu I)^2}}$, where $\mu$ is a
chemical potential. Diagonalized,  diagonal elements of $G$ become
$1$(-1) below(above) $\mu$. In the case of free Dirac fields,
$\mu$ is zero and all the negative energy states are fully filled
in the Dirac vacuum.

\section{Gaussian approximation and the state wavefunctionals}

When interactions between fermions exist, exact eigenfunctionals
are different from the above Gaussian. Variational method or
perturbation theory is applied in order to approximate true
eigenfunctionals. Although the variational method depends largely
on intuition in contrast to the systematic perturbative approach,
it could provide excellent results if trial states are  chosen
carefully.

In the Gaussian approximation, Gaussian wavefunctional with
variational parameters is used as a trial functional. We now
consider  the case of interacting fermion system where the
Hamiltonian has a general form,
\begin{eqnarray}
H = h_{i j}a_{i}^{\dagger}a_j +
v_{ijkl}a_{i}^{\dagger}a_{j}^{\dagger}a_k a_l .
\end{eqnarray}
Trial wavefunctional is chosen to be
\begin{eqnarray}
\Psi_G = e^{G_{ij}u_{i}^{*} u_j},
\end{eqnarray}
where $G$ is a variational parameter matrix. Then, the energy
expectation value is readily calculated with the aid of Eq.(5).
\begin{eqnarray}
E = \frac{\langle \Psi_G | H | \Psi_G \rangle}{\langle \Psi_G |
\Psi_G \rangle} = \frac{1}{2} (h + \Sigma)_{ij} \Omega_{ji}
-\frac{1}{4} \Sigma_{ij}  \Omega_{ji},
\end{eqnarray}
where $\Sigma_{ij}= ( v_{kijl}-v_{ikjl}) \Omega_{lk} $ and $
\Omega = (I+G) (G+\bar{G})^{-1} (I+\bar{G})$. Minimization of the
energy under a fixed number of particles is achieved through the
relation, $\frac{\partial }{\partial \bar{G}_{ij}}\left[ E -\mu
(\frac{1}{2} {\rm Tr} \Omega - N )\right]=0$, where $\mu$, the
chemical potential, is introduced as a Lagrange's multiplier.
Thus, the solution $G$ should satisfy the following condition;
\begin{eqnarray}
(I-{G}) \left( h+ \Sigma -\mu I\right)(I+{G})=0.
\end{eqnarray}
The equivalent condition for $\bar{G}$ is obtained by minimizing
with respect to ${G}$, and is given by $(I+\bar{G}) \left( h+
\Sigma -\mu I\right)(I-\bar{G})=0$ which provides the same $G$ as
in Eq.(11). As in Eq.(7), the nontrivial solution $G$ of Eq.(11)
is given by
\begin{eqnarray}
 G = -\frac{h+ \Sigma -\mu I}{\sqrt{(h+ \Sigma -\mu I)^2}} .
\end{eqnarray}
With this solution, the ground state energy $E_0$ is given as $E_0
= \frac{1}{2} {\rm Tr} (h+\Sigma) (I+G)$. The parent Hamiltonian
$\tilde{H}_0$ which has the above Gaussian vacuum is easily found
to be $\tilde{H}_0 = (h+\Sigma)_{i j}a_{i}^{\dagger}a_j$. In order
to obtain excitations based on the Gaussian wavefunctional, we
should have a unitary matrix $U$ which diagonalizes $G$ as $U G
U^\dagger = \tilde{I}$, where $\tilde{I}$ is a diagonal matrix
with elements of +1(-1) below(above) the  Fermi level $\mu$. Thus,
$G$ can be rewritten as follows;
\begin{eqnarray}
G= \left( \begin{array}{cc}
G_1 & G_2 \\
G_2^\dagger & G_4
\end{array} \right)
= \left( \begin{array}{cc}
U_1^\dagger & U_3^\dagger \\
U_2^\dagger & U_4^\dagger
\end{array} \right)
\left( \begin{array}{cc}
-I & 0 \\
 0 & I
\end{array} \right)
\left( \begin{array}{cc}
U_1 & U_2 \\
U_3 & U_4
\end{array} \right).
\end{eqnarray}
Therefore, we obtain an expression for $U$,
\begin{eqnarray} U = \frac{1}{\sqrt{2}}\left(
\begin{array}{ccc}
\sqrt{I-G_1 } &, & -\frac{1}{\sqrt{I-G_1 }}G_2  \\
\frac{1}{\sqrt{I+G_4 }}G_2^\dagger &, & \sqrt{I+G_4 }
\end{array} \right).
\end{eqnarray}
The Gaussian wavefunctional is diagonalized by the new Grassmann
variables $ \tilde{u}_i = U_{ij} u_j$ and $\tilde{u}_i^* = u_j^*
U_{ji}^\dagger$. We use capital(small) letters above(below) the
fermi see, so that
\begin{eqnarray}
\Psi_G = e^{ G_{ij}u_i^* u_j} = e^{\tilde{u}_a^* \tilde{u}_a
-\tilde{u}_{A}^* \tilde{u}_{A}} \ \ (a< k_F , \ A > k_F).
\end{eqnarray}
We define new creation and annihilation operators as follows;
\begin{eqnarray}
\tilde{a}_i^\dagger =\frac{1}{\sqrt{2}} \left( \tilde{u}_i^*
+\frac{\partial}{\partial \tilde{u}_i } \right) \ \ , \ \ \
\tilde{a}_i = \frac{1}{\sqrt{2}} \left( \tilde{u}_i
+\frac{\partial}{\partial \tilde{u}_i^* } \right) .
\end{eqnarray}
Excited wavefunctionals and their duals are obtained by these
operations to the Gaussian as
\begin{eqnarray}
\Psi_{\rm excited} &=&  \tilde{u}_{A_1}^{*} \cdot \cdot \cdot
\tilde{u}_{A_n}^{*}\tilde{u}_{a_1} \cdot \cdot \cdot
\tilde{u}_{a_n} e^{ G_{ij}u_i^* u_j} , \\
\bar{\Psi}_{\rm excited} &=&  \tilde{u}_{a_n}^* \cdot \cdot \cdot
\tilde{u}_{a_1}^* \tilde{u}_{A_n} \cdot \cdot \cdot
\tilde{u}_{A_1} e^{ G_{ij}u_i^* u_j} ,
\end{eqnarray}
which has an excitation energy of $\tilde{\epsilon}_{A_1}+\cdot
\cdot \cdot +\tilde{\epsilon}_{A_n} -(\tilde{\epsilon}_{a_1}+\cdot
\cdot \cdot +\tilde{\epsilon}_{a_n})$ where $\tilde{\epsilon}_i$
is the $i$th matrix element of $U(h+\Sigma)U^\dagger$. We note
here that any physical wavefunctional in the FJR could be
represented by multiplying only some combinations of $
\tilde{u}_A^*$ and $\tilde{u}_a $ to the Gaussian, while in other
representations, wavefunctionals should also contain Grassmann
derivatives for describing hole states and, thus, are quite
complicated in general. An important aspect of the present result
is that  Eqs.(17) and (18) have same structures to those in bose
systems [1].

\section{Rayleigh-Schr\"{o}dinger-Goldstone Perturbation Formalism beyond the Gaussian approximation}

In order to go beyond the Gaussian approximation by a perturbative
method based on the parent Hamiltonian $\tilde{H}_0$, Hamiltonian
is rearranged as
\begin{eqnarray}
H &=& \tilde{H}_0 + (H -\tilde{H}_0) \nonumber \\
&=& (h+\Sigma)_{i
j}a_{i}^{\dagger}a_j + \left(
V_{ijkl}a_{i}^{\dagger}a_{j}^{\dagger}a_k a_l - \Sigma_{i
j}a_{i}^{\dagger}a_j \right) \nonumber \\
&=&(\tilde{h}+\tilde{\Sigma})_{ij} \tilde{a}_i^\dagger
\tilde{a}_j +   \left( \tilde{V}_{ijkl}
\tilde{a}_i^\dagger \tilde{a}_j^\dagger
\tilde{a}_k \tilde{a}_l  -   \tilde{\Sigma}_{ij}
\tilde{a}_i^\dagger \tilde{a}_j  \right),
\end{eqnarray}
where $\tilde{h}=U h U^\dagger$,  $\tilde{\Sigma}=U \Sigma
U^\dagger$ and $\tilde{V}_{ijkl} =  U_{il} U_{jm}V_{lmpq}
U_{pk}^{\dagger} U_{ql}^{\dagger}$. Here, we note that the whole
formalism simply reduces to the conventional perturbation if
$\Sigma$ is set to zero. In the previous section, we have already
obtained whole spectrums of eigenvalues and eigenfunctionals of
the parent Hamiltonian $\tilde{H}_0$. Therefore, the
Rayleigh-Schr\"{o}dinger perturbation procedure [1] can be readily
adopted for improving the Gaussian approximation. However, in
order to carry out the calculation, it is necessary to have a
general rule for combination of Grassmann operations which is
similar to  Wick's theorem of the Green function approach. In the
following, we show that another perturbative approach in the time
independent formulation, namely Goldstone's expansion [27],
becomes an extremely useful tool in the present approach. Since
final results from Goldstone's expansion are same as from the
Rayleigh-Schr\"{o}dinger scheme order by order, we follow the
Goldstone's approach  to obtain the expressions for the
Rayleigh-Schr\"{o}dinger perturbation.

Goldstone's expansion is as follows. If $H=H_0 + H_1$,  $ H_0 |
\Phi_0 \rangle = E_0 | \Phi_0 \rangle$ and $ H | \Psi_0 \rangle =
(E_0 + \Delta E ) | \Psi_0 \rangle$, then
\begin{eqnarray}
| \Psi_0 \rangle = \sum_{n} {}^{'} \left( \frac{1}{E_0 -H_0 } H_1
\right)^n |
\Phi_0 \rangle _L    ,
\end{eqnarray}
\begin{eqnarray}
\Delta E =\sum_{n} {}^{'} \langle \Phi_0 |   H_1  \left( \frac{1}
{E_0 -H_0 } H_1 \right)^n  | \Phi_0 \rangle_C ,
\end{eqnarray}
where $| \Psi_0 \rangle$ is normalized as $\langle \Phi_0 | \Psi_0
\rangle =1$. The prime  in summation represents exclusion of terms
with zero denominators and subscripts $L$ and $C$ represent
$linked$ and $connected$ diagrams respectively. The $connected$
has the same meaning as in other many-particle theories. However,
the $linked$ diagrams in the Goldstone's usage are such that even
when some diagrams have unconnected parts, if unconnected parts
all have external lines, then, the diagrams are treated $linked$.

Let's expand the exact ground state wavefunctional $\Psi$ in terms
of the renormalized interaction, $H-\tilde{H}_0$. It gives $\Psi =
\sum_n \Psi_n$, where $\Psi_0 = \Psi_G = e^{G_{ij} u_i^* u_j}$.
Then, the $n$th order wavefunctional is given by
\begin{eqnarray}
\Psi_n = \sum {}^{'} \left( \frac{1}{E_0 -\tilde{H}_0 } (H-\tilde{H}_0)
\right)_L^n  e^{G_{ij} u_i^* u_j}   ,
\end{eqnarray}
where $\int D[u^* u] \bar{\Psi}_G \Psi$ = $\int D[u^* u] \bar{\Psi}_G \Psi_G$
is satisfied.

In order to represent explicitly the Grassmann wavefunctional, we
use the following useful relations. When capital(small) letters
are defined for states above(below) the fermi see, that is, $A, B,
\cdot \cdot \cdot > k_F$ and $a, b, \cdot \cdot \cdot < k_F$, we
arrive at simple commuting relations as follows,
\begin{eqnarray}
\tilde{a}_{A}^{\dagger} \Psi_G &=& \Psi_G \sqrt{2} \tilde{u}_{A}^{*} , \nonumber
\\
\tilde{a}_{A} \Psi_G &=& \Psi_G \frac{1}{\sqrt{2}} \frac{\partial}{\partial \tilde{u}_{A}^{*}}    , \nonumber
\\
\tilde{a}_{a}^{\dagger} \Psi_G &=& \Psi_G \frac{1}{\sqrt{2}} \frac{\partial}{\partial \tilde{u}_{a}}    , \nonumber
\\
\tilde{a}_{a} \Psi_G &=& \Psi_G \sqrt{2} \tilde{u}_{a}.
\end{eqnarray}
These commuting relations will be shown to reduce calculational
efforts substantially in the FJR. We further define $\bar{i}$ and
$i$ for notational simplicity as
\begin{eqnarray}
\tilde{a}_{i}^{\dagger} \Psi_G &=& \Psi_G \bar{i} , \nonumber \\
\tilde{a}_i \Psi_G &=& \Psi_G i  ,
\end{eqnarray}
where  explicit forms of  $\bar{i}$ and $i$ in the right-hand side
are defined through  Eq.(23) according to their position in the
fermi see. Then, we obtain
\begin{eqnarray}
\Psi_n =  \sum {}^{'} e^{G_{ij} u_i^* u_j} \left( \frac{1}{E_0 -\tilde{H}_0 }
\left( \tilde{V}_{ijkl}( \bar{i} \bar{j} k l ) -\tilde{\Sigma}_{ij}
 ( \bar{i} j ) \right) \right)_{L}^{n}.
\end{eqnarray}
The right-hand side above consists of combinations of  Grassmann
variables and Grassmann derivatives. We can easily see that if
there is no $u_i $ ($u_i^*$) on the right-hand side of
$\frac{\partial}{\partial u_{i}}$ ($\frac{\partial}{\partial
u_{i}^*}$), it should be zero. $Contraction$ is defined as a
Grassmann derivative result which is expressed by
\begin{eqnarray}
&&\begin{picture}(50,16) \put(5,16){\line(0,-1){5}}
\put(5,16){\line(1,0){7}} \put(12,16){\line(0,-1){5}} \put(0,0){ $
\bar{i}j = \frac{\partial}{\partial u_i} u_j \Theta(k_F - i) =
\delta_{ij} \Theta(k_F - i) $, }
\end{picture}   \\
&&\begin{picture}(50,16)
 \put(5,16){\line(0,-1){5}}
 \put(5,16){\line(1,0){7}}
 \put(12,16){\line(0,-1){5}}
 \put(0,0){ $ i \bar{j} = \frac{\partial}{\partial u_i^* } u_j^* \Theta(i- k_F) = \delta_{ij} \Theta(i - k_F) $. }
 \end{picture} \hspace{5cm}
 \end{eqnarray}
Self-contractions of bare interaction, $\tilde{V}_{ijkl}( \bar{i}
\bar{j} k l )$   are canceled out by $\tilde{\Sigma}_{ij}( \bar{i}
j )  $      because $\tilde{\Sigma}_{ij}( \bar{i} j ) =
(\bar{V}_{kijl}- \bar{V}_{ikjl}) ( U \Omega U^\dagger )_{lk}(
\bar{i} j )= 2(\bar{V}_{kijl}-\bar{V}_{ikjl})\delta_{lk}
(\Theta(k_F -l) - \Theta(l-k_F))( \bar{i} j )$. Therefore, the
interaction term can be simply expressed as
\begin{eqnarray}
\tilde{V}_{ijkl}( \bar{i} \bar{j} k l ) -\tilde{\Sigma}_{ij}
 ( \bar{i} j ) \equiv
\tilde{V}_{ijkl}( \bar{i} \bar{j} k l )'   ,
\end{eqnarray}
where the prime means that contractions should be performed with elements
outside $( \bar{i} \bar{j} k l )$.  With this prime
notation, the wavefunctional is expressed as
\begin{eqnarray}
\Psi_n &=& \left[ e^{G_{ij} u_i^* u_j} \sum {}^{'}
\frac{-\tilde{V}_{i_1 j_1 k_1 l_1}}{\delta E_1}
\frac{-\tilde{V}_{i_2 j_2 k_2 l_2}}{\delta E_2} \cdot \cdot \cdot
\frac{-\tilde{V}_{i_n j_n k_n l_n}}{\delta E_n} \right]  \nonumber
\\        && \times \left[
\left( ( \bar{i}_1 \bar{j}_1 k_1 l_1 )' ( \bar{i}_2 \bar{j}_2 k_2
l_2 )'   \cdot \cdot \cdot ( \bar{i}_n \bar{j}_n k_n l_n )'
\right)_L \right]     ,
\end{eqnarray}
where $\delta E_\mu \equiv \sum_{a=\mu}^n (\tilde{\epsilon}_{i_a}+
\tilde{\epsilon}_{j_a} - \tilde{\epsilon}_{k_a}-
\tilde{\epsilon}_{l_a}) $  means excitation energy. Each index in
the second square bracket in  Eq.(29) could be a Grassmann
variable or a Grassmann derivative. Therefore, in order to be
non-zero, all derivatives should be contracted out with
corresponding variables and finally remaining elements should be
Grassmann variables only. They are summarized as follows: (i) The
following steps are repeated for each $m=[0, 2n-1]$. (ii) A
possible set of $m$ pairs of bar and non-bar indices is selected.
Self-contractions are forbidden and  isolated parts fully
connected by contractions are also excluded. (iii) Each
contraction has the value given by Eq.(26) or Eq.(27). If odd
numbers of permutations are performed for contraction, (-1) is
multiplied. Remaining indices without contraction are converted by
$\bar{i} \rightarrow \sqrt{2} u_i^* \Theta(i-k_F)$ and $i
\rightarrow \sqrt{2} u_i \Theta(k_F-i)$. (iv) All other possible
$m$ contractions are performed.     Then, $\Psi_n$ is calculated
by multiplying the first square bracket term of Eq.(29) to each
contraction term and finally by summing out all indices.

The corresponding $n$th order energy ${\cal E}_n \ (\Delta E =
\sum_n {\cal E}_n)$ is calculated as
\begin{eqnarray}
{\cal E}_n &=& \frac{\int D[u^* u] \bar{\Psi}_G (H-\tilde{H}_0) \Psi_n }{ \int D[u^* u] \bar{\Psi}_G \Psi_G } \nonumber \\
&=& \sum {}^{'} {\tilde{V}_{i_1 j_1 k_1 l_1}}
\frac{-\tilde{V}_{i_2 j_2 k_2 l_2}}{\delta E_2} \cdot \cdot \cdot
\frac{-\tilde{V}_{i_n j_n k_n l_n}}{\delta E_n}  \nonumber
\\        && \times
\left( ( \bar{i}_1 \bar{j}_1 k_1 l_1 )' ( \bar{i}_2 \bar{j}_2 k_2 l_2 )'   \cdot \cdot \cdot
( \bar{i}_n \bar{j}_n k_n l_n )' \right)_C     ,
\end{eqnarray}
where subscript $C$ means that we should perform fully connected
contractions.

As well as energy values, Gaussian expectation values of any
operators like $ {\cal O}(\tilde{a}_i^\dagger , \tilde{a}_j ,
\cdot \cdot \cdot ) $ are also simply expressed in the FJR as
follows,
\begin{eqnarray}
\langle {\cal O} \rangle &=& \frac{\int D[u^* u] e^{G_{ij}u_i^*
u_j} {\cal O}(\tilde{a}_i^\dagger , \tilde{a}_j , \cdot \cdot
\cdot )  e^{G_{ij}u_i^* u_j}  }{\int D[u^* u] e^{2 G_{ij}u_i^*
u_j} } \nonumber \\
&=& \frac{\int D[u^* u] e^{2 G_{ij}u_i^* u_j} {\cal O}(\bar{i}, j,
\cdot \cdot \cdot )}{\int D[u^* u] e^{2 G_{ij}u_i^*
u_j} } \nonumber \\
&=& {\cal O}(\bar{i}, j, \cdot \cdot \cdot ) |_{u^{*}, u =0 } ,
\end{eqnarray}
where $G$ should be the variational solution, Eq.(12), and the
notational abbreviations, Eq.(16) and (24) are also used. The
third line of Eq.(31) is equivalent to the full contractions
allowing unconnected ones.

As examples, we calculate the first order wavefunctional $\Psi_1$
and the second order energy ${\cal E}_2$. If we use capital(small)
letters above(below) the Fermi see in the correction parts, they
are represented as
\begin{eqnarray}
\Psi_1 =-4 e^{G_{ij} u_i^* u_j} \sum_{A,B,a,b}
\frac{\tilde{V}_{ABab}}{\tilde{\epsilon}_{A} +\tilde{\epsilon}_{B}
-\tilde{\epsilon}_{a} -\tilde{\epsilon}_{b}} \tilde{u}_A^*
\tilde{u}_B^* \tilde{u}_a \tilde{u}_b ,
\end{eqnarray}
\begin{eqnarray}
{\cal E}_1 &=& 0 , \\
 {\cal E}_2 &=&
-2\frac{\tilde{V}_{abAB}(\tilde{V}_{ABab}-\tilde{V}_{BAab})}{\tilde{\epsilon}_{A}
+\tilde{\epsilon}_{B} -\tilde{\epsilon}_{a} -\tilde{\epsilon}_{b}}
.
\end{eqnarray}
Here, we note that ${\cal E}_1$ is always zero and $\Psi_1$ has no
terms like $\tilde{u}_A^* \tilde{u}_a$
 because self-contraction is forbidden.

 In the case of electron
gas, the quantum numbers are usual  wavevector $k$ and spin
$\sigma$ and the interaction has a form,
\begin{eqnarray}
 {V}_{ijkl} \rightarrow {V}_{ k_{1} \sigma_{1} , k_{2} \sigma_{2} ,
 k_{3} \sigma_{3} , k_{4} \sigma_{4} }  = \frac{1}{2} v( | k_4 - k_1 | )
\delta_{k_1 + k_2 , k_3 + k_4} \delta_{\sigma_1 \sigma_4}
\delta_{\sigma_2 \sigma_3} .
\end{eqnarray}
For a homogeneous case, the Gaussian variational solution $G$ of
Eq.(12) is already diagonalized by $k \sigma$ representation.
Therefore, we can use the Grassmann variables $u_{k \sigma}^* ,
u_{k \sigma}$ as basis states instead of $\tilde{u}_{k \sigma}^* ,
\tilde{u}_{k \sigma}$. The energy component of parent Hamiltonian
$\tilde{H}_0$ is calculated as $\tilde{\epsilon}_{k \sigma} = (h +
\Sigma)_{k \sigma, k \sigma}= \epsilon_k^0 - \sum_{q} v(q) n_{k+q
\sigma}$ where $\epsilon_k^0$ is the free electron energy band and
$n_{k \sigma}$ is the Fermi distribution function at zero
temperature. Thus, we finally have
\begin{eqnarray}
\hspace{-1cm} \Psi_1 =-2 e^{G_{k \sigma , k' \sigma'} u_{k
\sigma}^* u_{k' \sigma'}} \sum_{k k' q \sigma \sigma'}
\frac{v(q)}{\tilde{\epsilon}_{k'+q} +\tilde{\epsilon}_{k-q}
-\tilde{\epsilon}_{k} -\tilde{\epsilon}_{k'}} {u}_{k'+q \sigma}^*
{u}_{k-q \sigma'}^* {u}_{k \sigma'} {u}_{k' \sigma} , \ \ \ \
\end{eqnarray}
\begin{eqnarray}
{\cal E}_2 =&-& \sum_{k k' q } \frac{v(q)^2 n_k n_{k'}
(1-n_{k'+q})(1-n_{k-q})}{\tilde{\epsilon}_{k'+q}
+\tilde{\epsilon}_{k-q} -\tilde{\epsilon}_{k}
-\tilde{\epsilon}_{k'}} \nonumber \\
&+& \sum_{k k' q } \frac{v(q) v(q') n_{k+q+q'} (1- n_{k+q'}) n_k
(1-n_{k+q})}{\tilde{\epsilon}_{k+q} +\tilde{\epsilon}_{k+q'}
-\tilde{\epsilon}_{k} -\tilde{\epsilon}_{k+q+q'}} .
\end{eqnarray}
The first order wavefunctional has a very simple form and higher
order ones are also expected to have such simple forms in this
variational FJR. We find that the above ground state energy is
same as the zero temperature limit of the variational perturbation
theory [25] previously obtained from a functional integral
representation, because Gaussian approximations of both
representations employed same variational basis. For a numerial
energy value, we modeled an electron gas with a bare kinetic
energy $\epsilon_k^0 = \frac{\hbar^2 k^2}{2 m_e \alpha}$ and a
Yukawa-type interaction $v(q) = {4 \pi \gamma e^2}/{V ( q^2 + (
\frac{\lambda}{a_0} )^2 )}$ where $m_e$ and $e$ are the bare
electron mass and the charge, and $V$ and $a_0$ are the system
volume and the Bohr's radius respectively. $\alpha, \gamma$, and
$\lambda$ are parameters to modulate the mass, charge and
interaction range. In Fig.1, we plot
 the second order contribution to the ground state energy as a function of $r_s$, which is
the average distance ratio between electrons defined by $V=
\frac{4}{3} \pi (r_s a_0 )^3 N$. We find the second order
contribution is much reduced in the present variationl procedure
than the conventional one which usually gives exceedingly negative
result. This indicates much better convergence  for the
variational perturbation theory as expected.

\section{Summary}
We have presented a Rayleigh-Schr\"{o}dinger-Goldstone
perturbation formalism using the Floreanini-Jackiw representation
on fermion systems.  With the aid of Goldstone's expansion, formal
expressions for  the ground state wavefunctional and the
corresponding energy are obtained in terms of  a renormalized
interaction. It is shown that this formalism can be used for both
standard perturbation and variational perturbation schemes. Useful
commuting relations  between  creation and annihilation operators
with the Gaussian wavefunctional have been found. As examples, we
have calculated the first order wavefunctional and the sencond
order ground state energy of electron gas with Yukawa-type
interaction.

\ack
This work has been supported by the Korea Research Foundation
(KRF 2003-005-c00011). C.K.K. acknowledges support from the Third
World Academy of Sciences through visiting associateship to
Institute of Physics, Beijing, where part of this work was carried
out.

\Figures
\begin{figure}
\caption{The second order energy correction (Eq.(36)) per particle
for the electron gas with the Yukawa-type interaction. ($\alpha =
1.1$, $\gamma=1.1$ and $\lambda=0.1$) ; filled circles
   are from the present variational perturbation and open circles
   from
   the conventional perturbation.}
\end{figure}
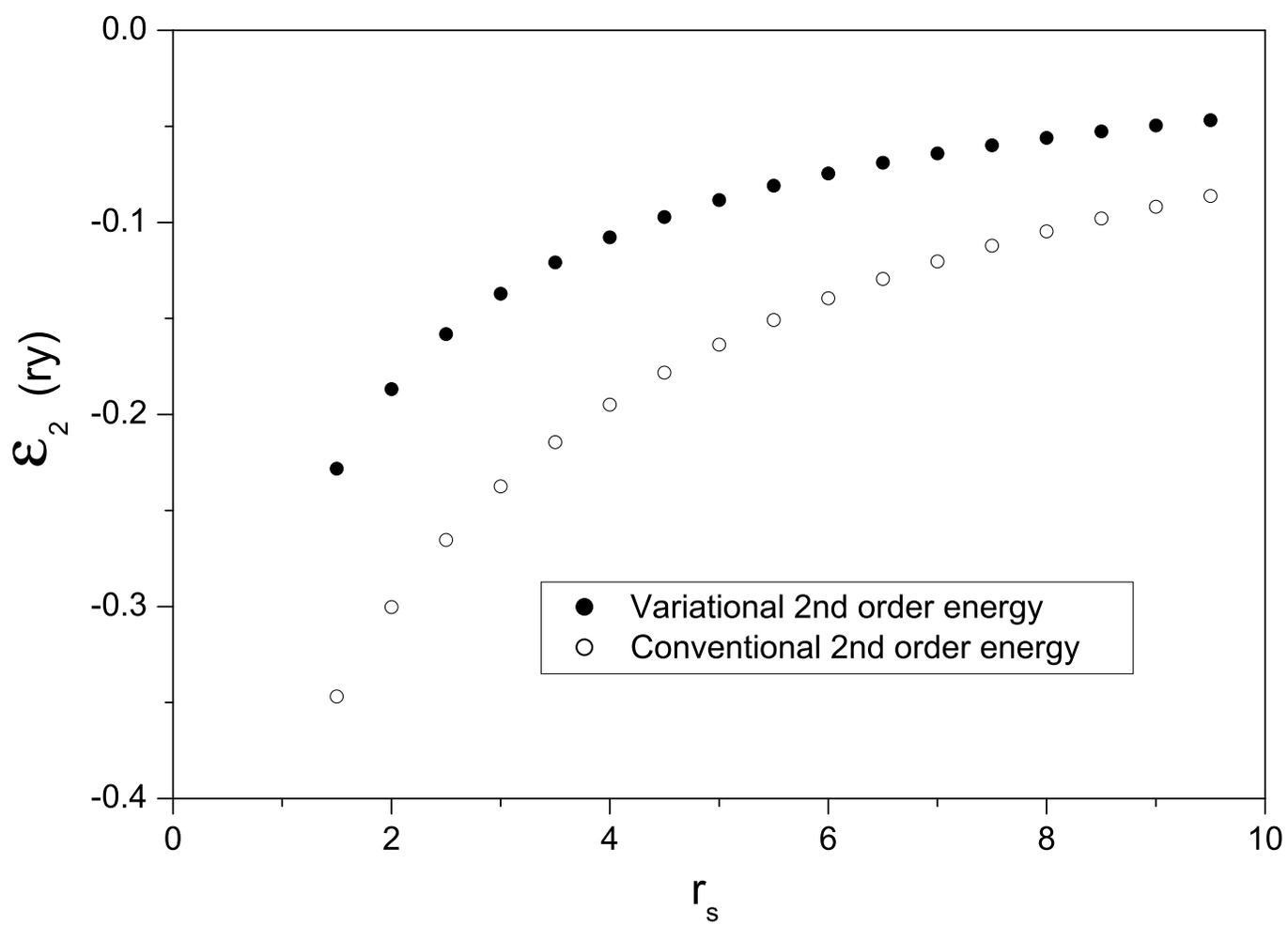
\end{document}